\documentclass{sig-alternate-10pt}

 \usepackage{multirow}
 \usepackage{url}
\usepackage{placeins}
\usepackage{tablefootnote}
\usepackage{amssymb,amsmath}
\usepackage{gensymb} 
\usepackage{graphicx}
\usepackage{caption}
\usepackage{subcaption}
\usepackage[numbers]{natbib}
\usepackage{array}
\usepackage{gensymb}
\usepackage[paperwidth=8.5in, paperheight=11in,top=1.5in, bottom=1.5in, left=1in, right=1in]{geometry}

\newlength\figwidth
\begin{document}

\title{The Web for Under-Powered Mobile Devices: Lessons learned from Google Glass }

%
%
%
%
%

\numberofauthors{3} 
%
\author{
%
%
\alignauthor
Jagmohan Chauhan\\
       \affaddr{UNSW \& NICTA}\\
\alignauthor
Mohamed Ali Kaafar\\
       \affaddr{NICTA}\\
\alignauthor
Anirban Mahanti \\
       \affaddr{NICTA}\\
    }

\maketitle
\begin{abstract}
This paper examines some of the potential challenges associated with enabling a seamless web experience on underpowered mobile devices such as Google Glass from the  perspective of web content providers,  device, and the network. We conducted experiments to study the impact of webpage complexity, individual web components and  different application layer  protocols while accessing webpages    on the performance of  Glass browser, by measuring webpage load time, temperature variation and power consumption and compare it to a smartphone. 
Our findings suggest that (a)  performance of Glass compared to a smartphone in terms of  power consumption and webpage load time deteriorates with increasing webpage complexity (b)  execution time for popular JavaScript benchmarks  is about 3-8 times higher on Glass compared to a smartphone, (c)  WebP is more energy efficient image format  than JPEG and PNG, and (d)  seven out of 50  websites studied  are  optimized for content delivery to Glass. 

\end{abstract}



\section{Introduction}
Over the past two decades, the web has changed significantly, evolving from  basic webpages with hyperlinks to a substantially more complex ecosystem with dynamic webpages and dynamic content delivery models. While the web ecosystem has become complex, the devices accessing the web are becoming smaller, moving from desktop to tablets to smartphone and now wearables. The trade-off between ever increasing web complexity and gradual miniaturization of devices having limited processing capabilities and power posits an interesting question: How does the current web ecosystem  impact the performance of underpowered devices? In this paper, we address this question by considering a popular smartglass: Google  Glass. 

    
The main focus of this paper is (a) to quantify the web browser performance on Glass and  compare it to a smartphone browser,  (b) to understand   the  factors  potentially responsible for the performance differences between the two devices, and (c) provide insights as to how content providers can offer a better user quality of experience on underpowered devices. A \textit{profiler} is developed for both Glass and smartphone to monitor four performance metrics:  power consumption, temperature variation, downloaded bytes and webpage load time while loading the webpages using WiFi over a series of experiments.  The initial experiment consists of accessing 50 popular    websites under various categories from Alexa Top 500.  The next set of experiments deals with loading synthetic webpages, executing popular JavaScript benchmarks and loading different image formats.   The final experiment is to access the websites measured in the initial experiment using HTTPS.  Additionally, we analyzed tcpdumps and examined  the  web objects served by  50 websites  to find if they  provide different web experience to Glass and smartphone.    

Based on our experiments, the important results are:
\begin{itemize}
\item  The performance of Glass compared to a smartphone in terms of total power consumption and webpage load time deteriorates with increasing number of  web objects,  number of servers accessed and number of JavaScripts on a webpage. The webpage load time  for the landing page of  popular websites is very high (2x on average) on Glass compared to a smartphone. However, 2x  higher webpage load time only results in  1.2x (on average) higher total power consumption on Glass due to a lower  rate of power consumption than a smartphone.
\item JavaScript is the most resource intensive web component. 
 The Glass browser is  about 3-8 times slower than the Chrome browser on a Nexus 5 smartphone,  while executing the same JavaScript benchmarks. The execution time for 3rd party analytics and ad scripts on Glass is about 2x that of smartphone.
\item WebP image format is  more energy efficient   than JPEG and PNG on Glass. For example, using WebP   instead of JPEG on \url{m.wikihow.com} results in 45 \% savings in power consumption and 50 \% lower   webpage load time.
\item The cost of accessing  a website using HTTPS on Glass when compared to a smartphone increases with increasing number of web objects, webpage size and the  number of servers accessed through the webpage. For instance, Glass power consumption is 27 \% lower than smartphone for webpages with less than 64 web objects. However, the    power consumption on Glass  becomes 17 \% higher to that of  smartphone when loading webpages having more than 64 web objects.
\item Seven out of the 50 studied  websites are optimized for content delivery to  Glass. For example, \url{m.espn.go.com} optimize  by serving images according to  the Glass screen dimensions. 

\end{itemize}

The rest of the paper is organised as follows. Section 2 discusses related work. 
Section 3 explains  experimental setup.  Performance results are discussed in Section 4. Section 5 concludes the paper. 

\section{Related Work}
Prior works  have investigated the performance  of web browser on desktop  and smartphones  from different perspectives such as caching, protocols and webpage content \cite{Xiao,web}, webpage complexity \cite{micheal} and energy \cite{Thiagarajan}.  In contrast, our major focus is, (a) to find out the web browsing  performance issues on an underpowered device such as Glass and, (b)  how today's webpage complexity and design   contribute to such issues.

Xiao et al. \cite{Xiao} developed WProf to  profile web browser activities on a desktop to measure the webpage load time accurately. Their study on 350 webpages  concluded   that synchronous JavaScript contributes significantly to the webpage load time and  SPDY  \cite{misc15} reduces webpage load time only for low bandwidth networks. SPDY is  a network protocol  developed  at Google to achieve smaller web page load latencies. Michael et al. \cite{micheal} demonstrated the performance impact of  increasing webpage complexity  on webpage load time  by characterizing   2000 websites landing pages on a desktop browser. Similar to earlier observations by Michael et al. \cite{micheal}, we found out that the    number of requested web objects, number of servers and number of javascripts   have the highest impact on the webpage load time.   Our work differentiates from the aforementioned two studies  by  studying  web browser performance not only in terms of webpage loading time but also power consumption and temperature variation on mobile devices. 

Thiagarajan et al. \cite{Thiagarajan} created a system to exclusively measure smartphone browser energy for popular websites based on their content.  In contrast, our work measures browser performance along multiple metrics taking into account    webpage complexity, webpage content and different application protocols.   Qian et al. \cite{web} suggested  factors including protocol overhead, webpage content and caching  affect resource utilization for mobile web browsing. We measured  the impact of similar factors on Glass and smartphone browsers.

\section{Experimental Setup}
The experimental setup  consists of a Google Glass, a Nexus 5 smartphone (Chrome browser 42.0) and a laptop. We also repeated the same set of experiments with Sasmung Galaxy S4 and Nexus 5X, and  observed performance difference trends  between Glass and the two smartphones to be similar to Nexus 5. For brevity, we only report results  for  Nexus 5. The laptop runs Mac OS X, has 8 GB of RAM and 2.6 GHz processor.  Our experiments can be broadly divided into two categories: accessing real webpages on the Internet from Glass and smartphone, and accessing synthetic webpages hosted on a local Apache web server running on a laptop  from Glass using  WiFi.   The synthetic webpages are created for two scenarios: (a) Hosting  landing pages of websites that are considered for studying the impact of various web components on browser, and (b) Image format comparison experiments.  Hosting webpages on a local server provides a way to control webpage content according to experimental needs. More details are presented  in Section 4.2.1  and 4.2.3.   Overall goals of our experiments is to study the impact of webpage complexity, individual web components and accessing webpages using different application layer protocols  on the performance of Glass and smartphone browser and compare them.  

We created  an application for both Glass and smartphone, which invokes the browser  with the  website url to be accessed as an input.  Browser cache is emptied before each experiment. We also developed a \textit{profiler}   app for both the devices. \textit{Profiler}  runs in the background and collects  the following performance metrics every second and writes them to a file with timing information for later analysis:
\begin{itemize}
\item Power consumption is obtained by multiplying the current and voltage readings obtained from \textit{current\_now} and \textit{voltage\_now} files  present at  \textit{/sys/class/power\_supply/battery}. 

\item Glass device temperature  is obtained by reading \textit{/sys/devices/platform/notle\_pcb\_sensor.0/\\temperature} file.

\item Downloaded bytes is obtained by using \textit{getUidRxBytes} function from Android framework \cite{sdk}. 
 \item  Webpage load time is measured from the time of first DNS request (start time) to the time when browser receives the last web object (end time). The \textit{profiler} extracts  the start and end time from the  browser logs.
\end{itemize}
 


All the experiments are repeated six times and the results reported are averages unless stated otherwise. 
Each experiment is first performed on Glass  and then  on smartphone. The battery of the devices  is fully charged. Glass is always allowed to cool down to  around 37 \degree C before  commencing experimentation.    While taking measurements,  only \textit{profiler}, app to invoke the  browser and the   browser are running on the devices. 

\begin{figure*}[!ht]
\centering
\begin{subfigure}{\figwidth}
     \includegraphics[height=0.25\textheight]{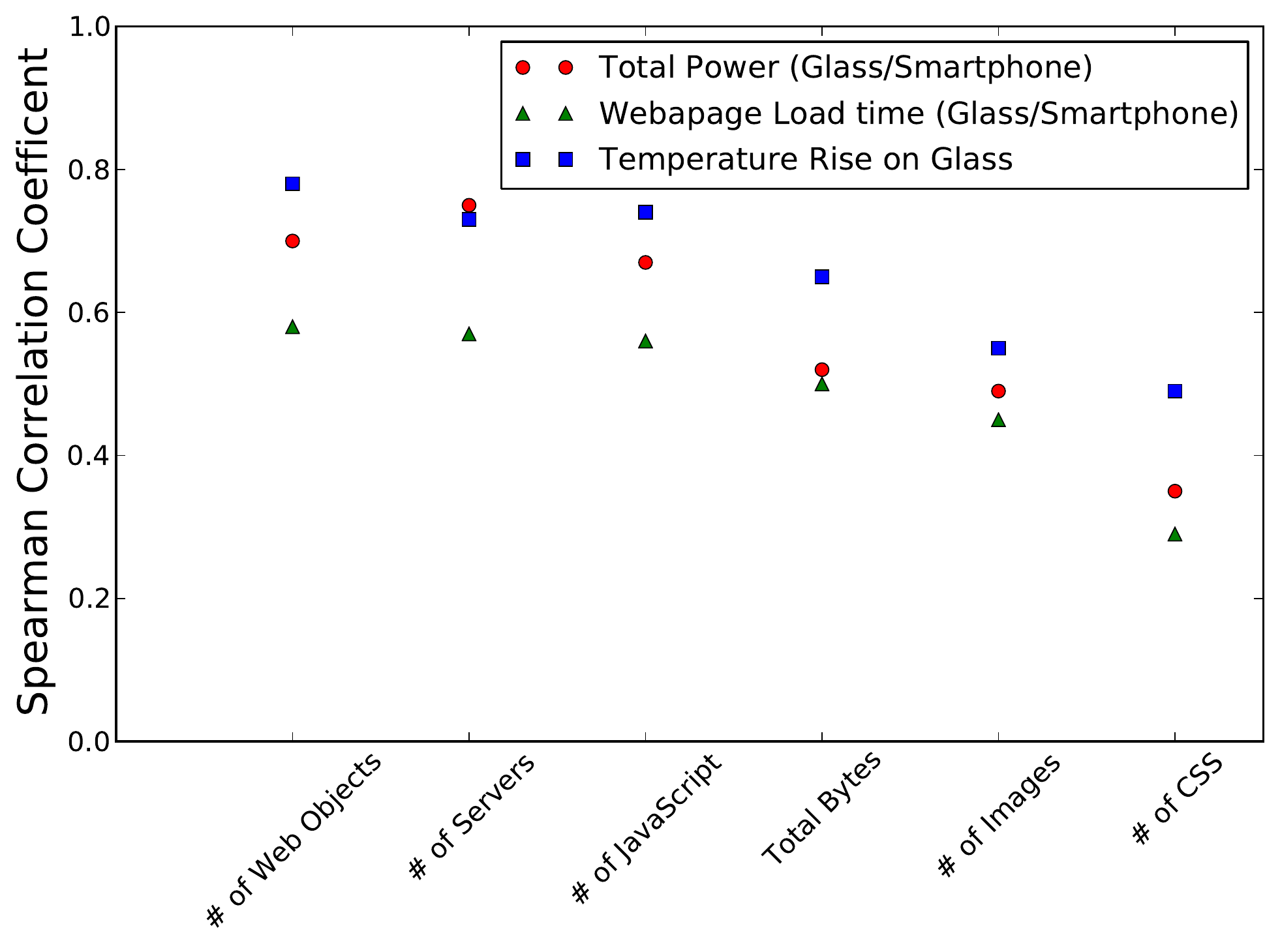}
     \caption{Correlation Analysis}
     \label{fig-corrfig}
\end{subfigure}%
\begin{subfigure}{\figwidth}
     \includegraphics[height=0.25\textheight]{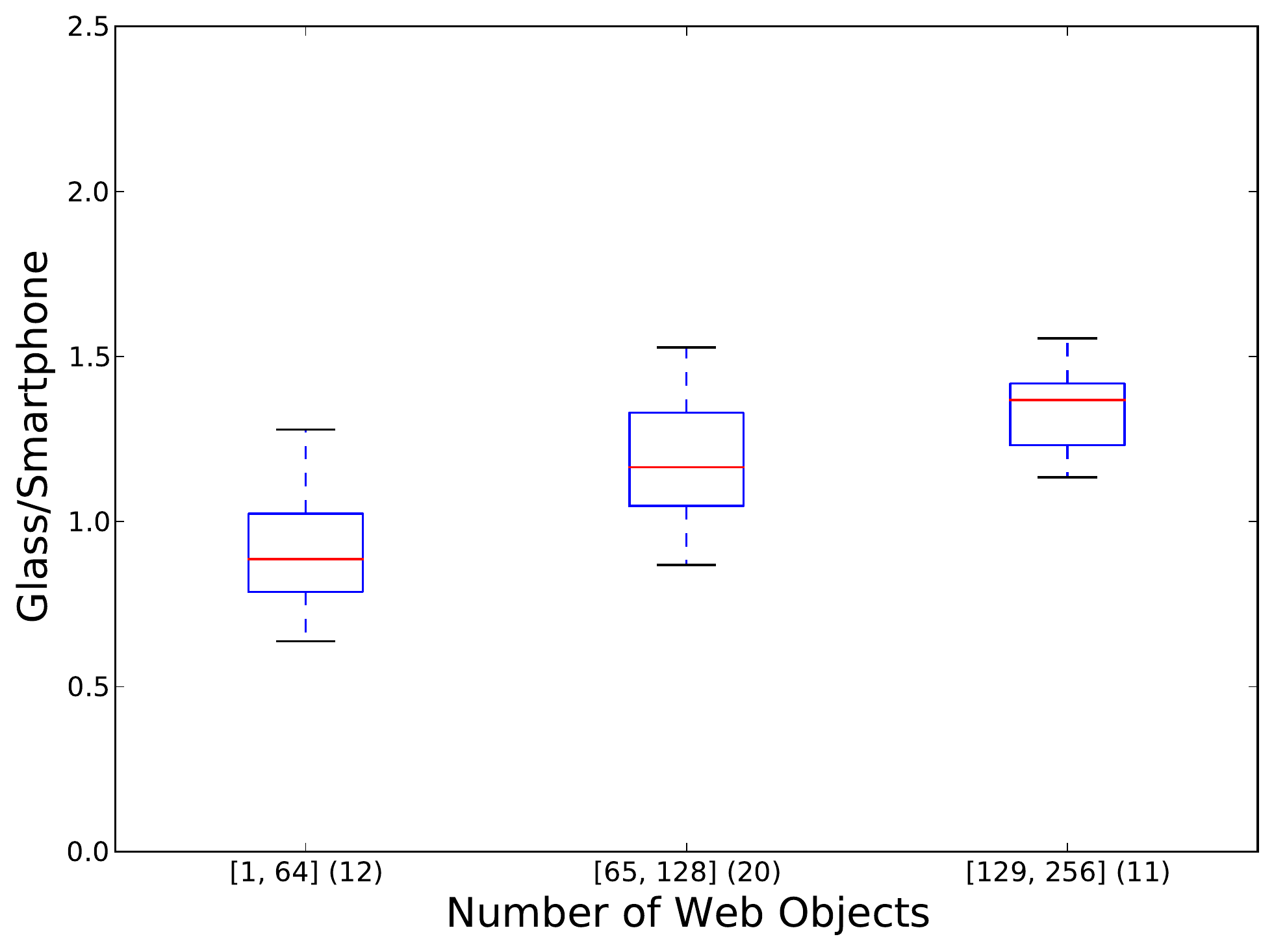}
     \caption{Total Power Consumption (mW)}
     \label{fig-boxplot1}
\end{subfigure}
\begin{subfigure}{\figwidth}
	\includegraphics[height=0.25\textheight]{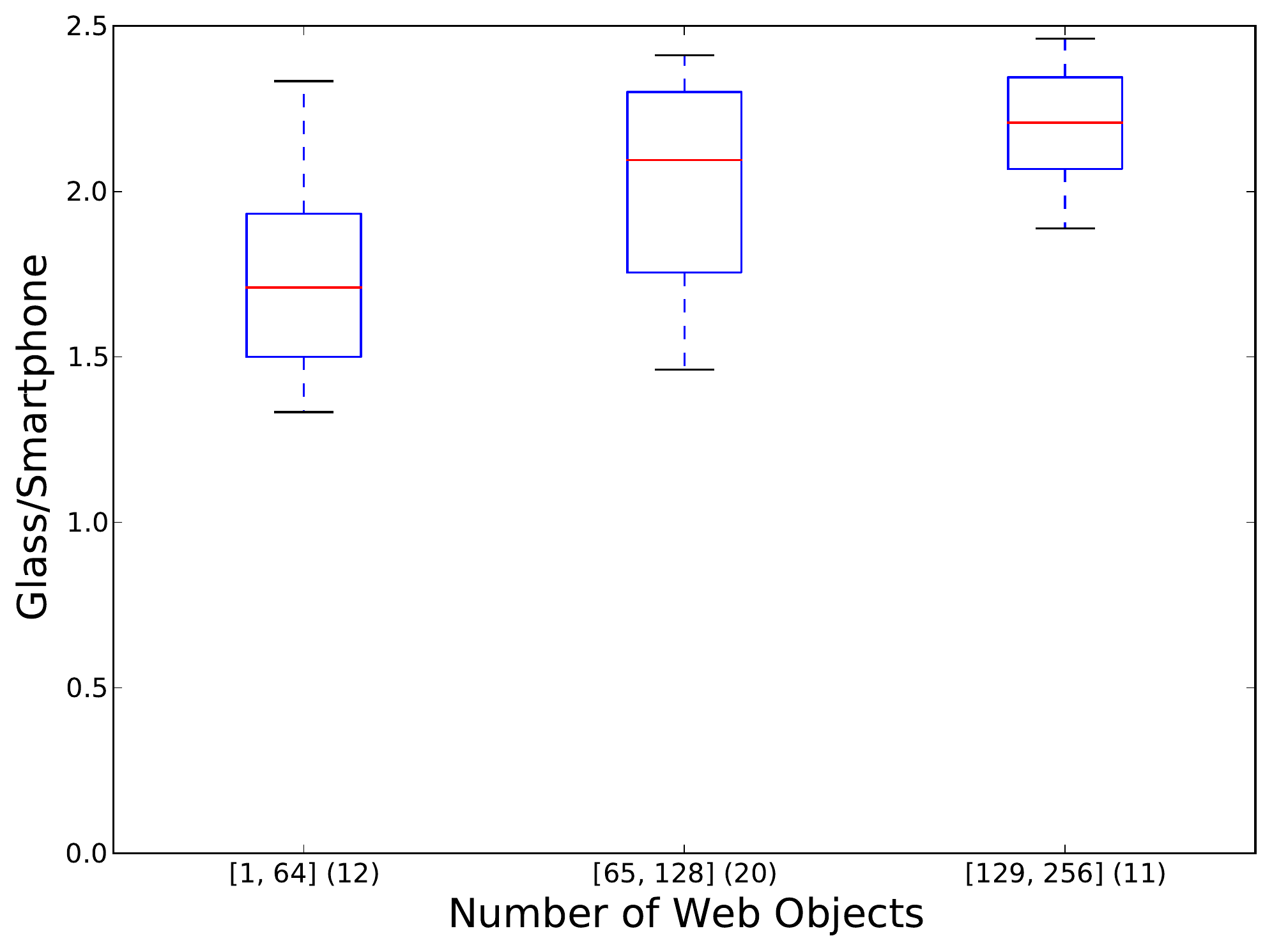}
    \caption{Webpage Load Time}
    \label{fig-boxplot2}
\end{subfigure}%
\begin{subfigure}{\figwidth}
     \includegraphics[height=0.25\textheight]{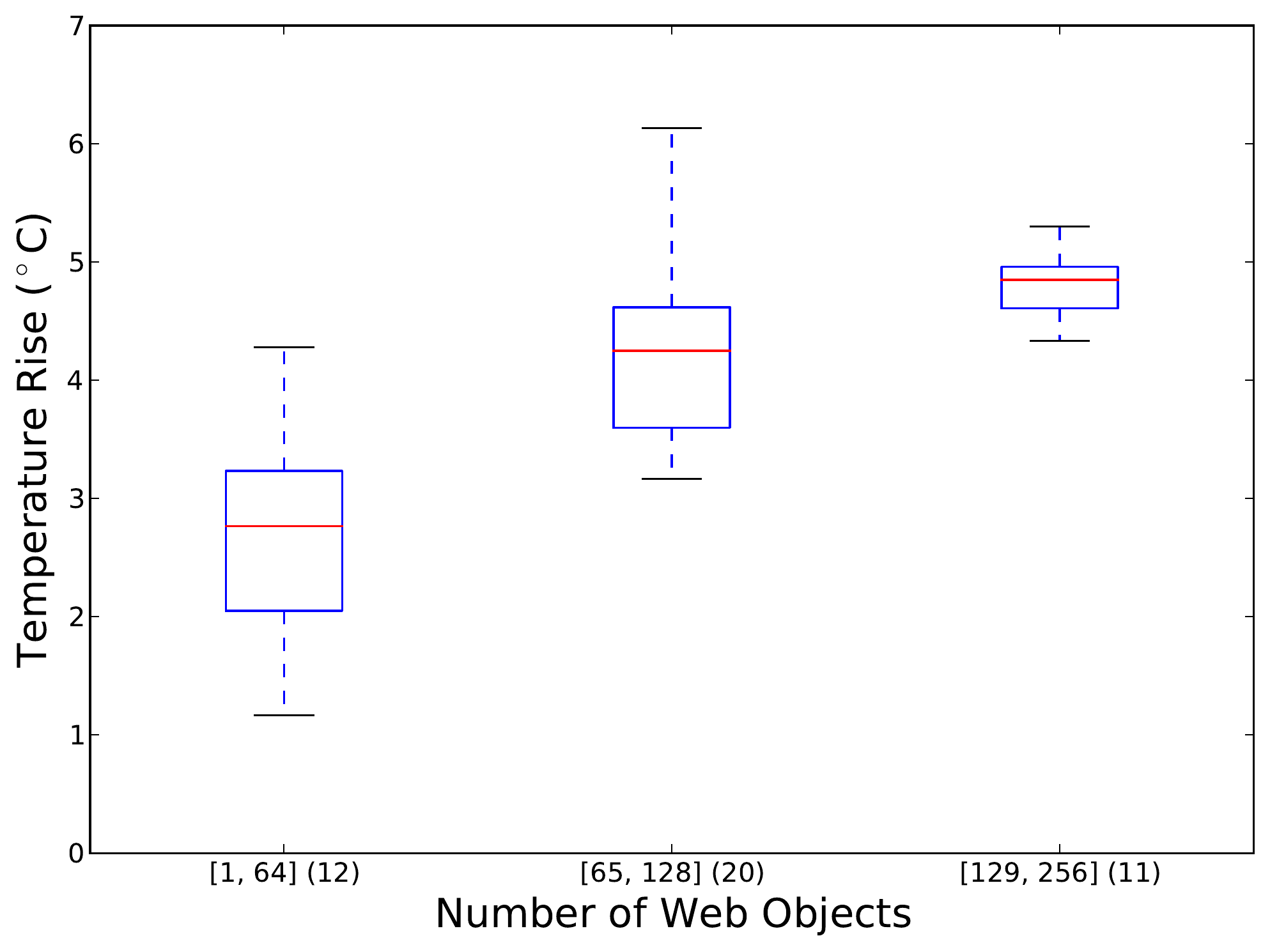}
     \caption{Temperature Rise on Glass}
     \label{fig-boxplot3}
     \end{subfigure}
\caption{Comparison of Web Browsing Performance}
\label{fig-browserperf}
\end{figure*}

\section{Results}

\subsection{Browser Performance for Popular Websites}

Our first experiment studies how the complexity of webpages affects the  performance of  Glass in comparison to a smartphone. The top categories of websites being accessed from Glass are media, entertainment, sports, news, informational and technology \cite{misc14}. We picked 50 popular websites (mentioned in  Appendix) from Alexa   Top 500 under the aforementioned  categories.  We found  seven websites  to be  optimized for content delivery on Glass and are hence discussed separately in Section 4.4.

\begin{figure*}[ht]
  \centering
  \begin{subfigure}{0.8\figwidth}

    \includegraphics[width=\textwidth]{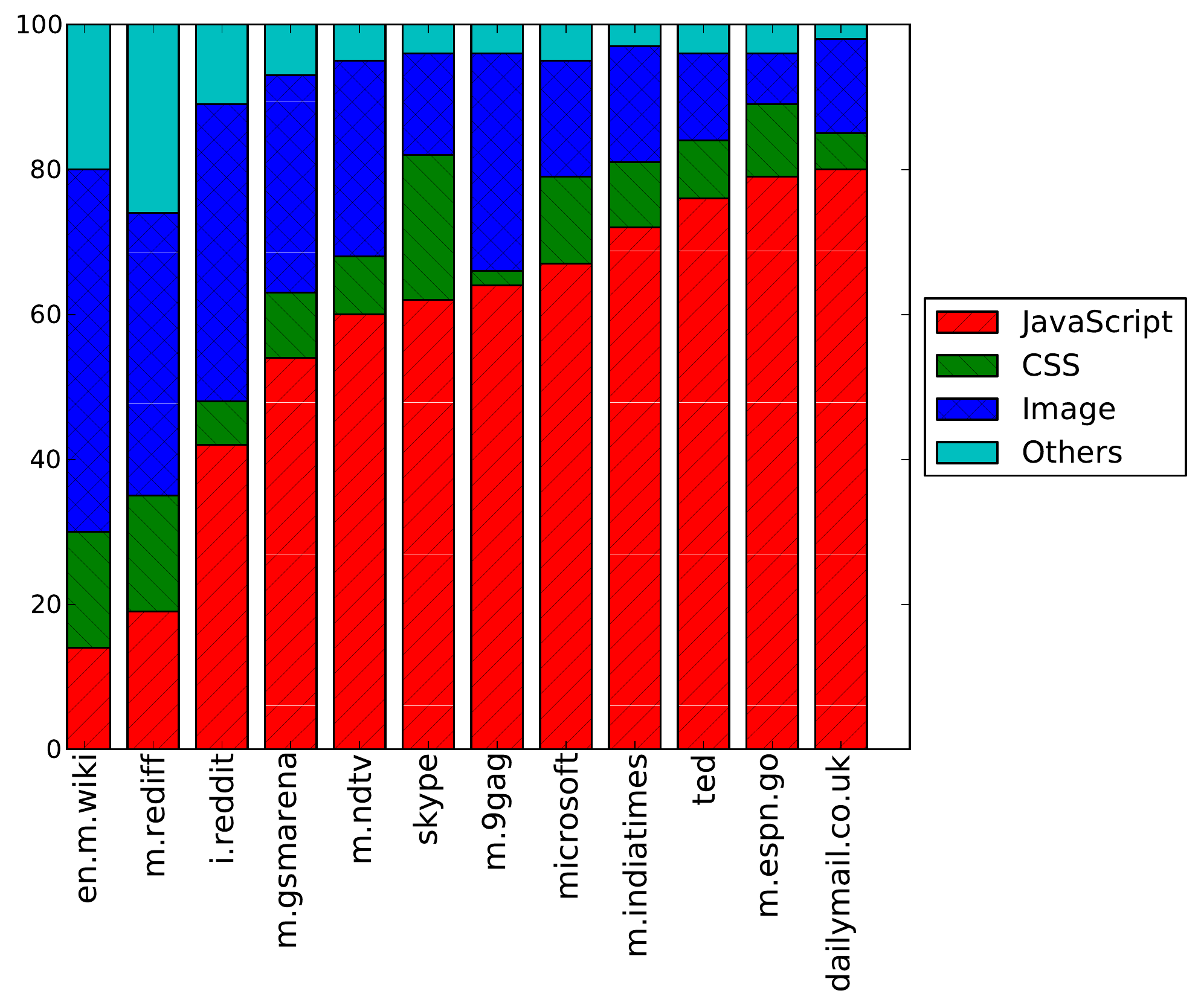}
\caption{Percent of total power consumption by web components. ``Others"  includes text and fonts.}
\label{fig-webcomp}
\end{subfigure}
  \begin{subfigure}{0.8\figwidth}
    \includegraphics[width=\textwidth]{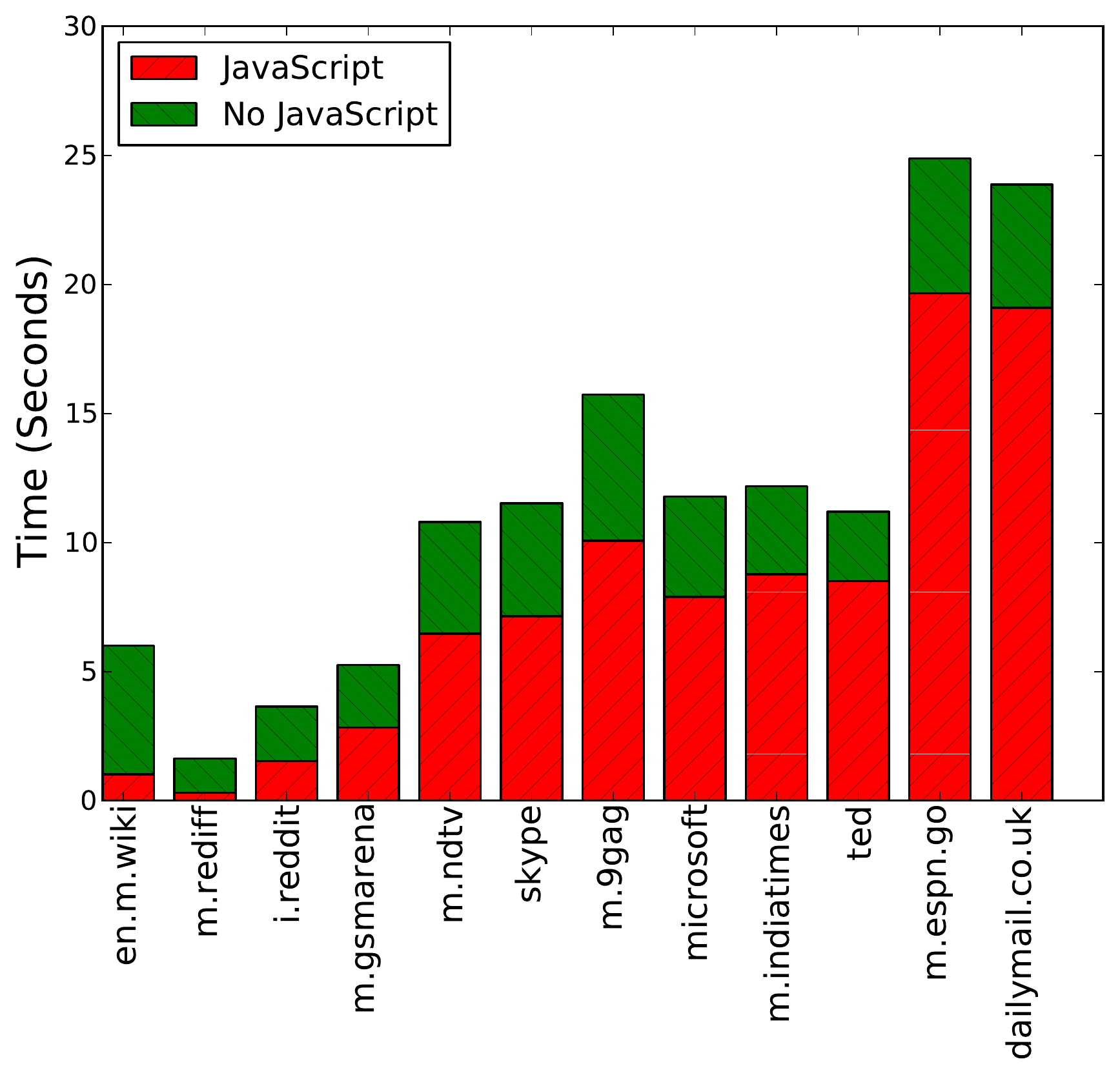}
\caption{Contribution to Webpage Load   Time by  JavaScripts compared to all other web components (CSS+Images+Text+Fonts)  }
\label{fig-jstiming}
\end{subfigure}
\caption{Glass Browser Performance for Synthetic Landing Webpage of Websites}
    \label{fig-syn}

\end{figure*}

To compare  the performance  between  Glass and a smartphone, we choose to represent relative performance (Glass/Smartphone) ratio  in the results for each metric: total power consumption and webpage load time. Temperature variation metric is only shown for Glass. A ratio of less than one means Glass fares better than smartphone and vice versa.  To  understand the causes for performance differences between Glass and smartphone, we first did a correlation analysis for three aforementioned performance metrics  against  webpage complexity factors: number of web objects, number of servers contacted, total bytes, number of JavaScripts, number of CSS and number of images.  Next, websites are binned based on the common highest correlated factor amongst all the factors to depict the relative performance. The results are shown in Figure \ref{fig-browserperf}. 

Figure \ref{fig-corrfig} shows that across all three performance metrics, the three most correlated web complexity  factors are the total number of web objects, the number of Javascripts within these objects, and the number of servers contacted while loading the webpage with a very high correlation coefficient (0.6-0.8).  As Glass is computationally less powerful than  a smartphone, the performance gap increases with every connection a browser has to create to fetch a web component, which is determined by the number of web objects and the number of servers to be accessed. Amongst the web components, the highest impact   is caused  by  JavaScripts,  shown in the next section.  



 Figure \ref{fig-boxplot1}-\ref{fig-boxplot3}   confirms the strong correlation between three performance metrics and the number of web objects.  On the x-axis, the first item is the   interval representing the number of web objects. The second item refers to the number of websites in the interval.  The   total power consumption and webpage load time for Glass in comparison to a smartphone deteriorates with  the increasing number of  web objects. The temperature rise on Glass also shows the same observation. As, Glass and Nexus 5 are using same WiFi technology and  access point, we suspect the processor to be  mainly responsible for the deterioration. Glass  processor capability (1.2 GHz)  is almost half  that of Nexus 5 (2.25 GHz), which results in slower processing of web content and hence prolong webpage load time. Our results confirms the webpage load time for Glass to be roughly two times that of Nexus 5. The  cost of accessing websites can be reduced  on Glass by designing webpages  having a small number of web objects (fewer JavaScripts) being fetched from fewer  servers. 

\subsection{Browser Performance for Web Components}
To understand how the  modern website design affects the Glass browser performance, we performed three different experiments. First, we measured the webpage load  time and power consumption for three key  elements: CSS, JavaScript and Images for the synthetic landing webpage of websites on Glass. Second, we measure the execution time of  popular JavaScript benchmarks, analytics and ad scripts on Glass and smartphone.  Third, we study the power consumption of JPEG, PNG and WebP image formats on Glass.

\subsubsection{Synthetic WebPages: Breakdown Analysis by Web Components}

To conduct our experiments, we created a copy of the website landing pages on a local Apache web server (version 2.2.29).  The local HTML copy  allows us to systematically add or remove web components. The local server only contains a copy of website landing HTML page. All the web objects embedded inside the webpage are still served by the original servers. We chose to do so because   modern websites are complex and have dynamic content, which makes it impractical to store each and every web object embedded inside the  landing page on the local server. 

The  energy consumed by  each web component is estimated by comparing   the energy consumption used for loading the entire webpage to the energy consumption needed for loading the webpage with a specific type of web component removed by filtering it out from the HTML code. 
 We followed two criteria  while selecting websites for these experiments: (a) the website landing HTML page should not make any web object request to the local server, (b) images present on the landing page should not be embedded  inside a CSS or JavaScript. These two   criteria prevents the skewness in measurements and ensures that  local server only gets a  request for the landing  HTML webpage. 12 out of 50 websites passed both  the criteria.

\begin{figure*}[ht]
  \centering
     \begin{subfigure}{0.7\figwidth}

  \includegraphics[width=\textwidth]{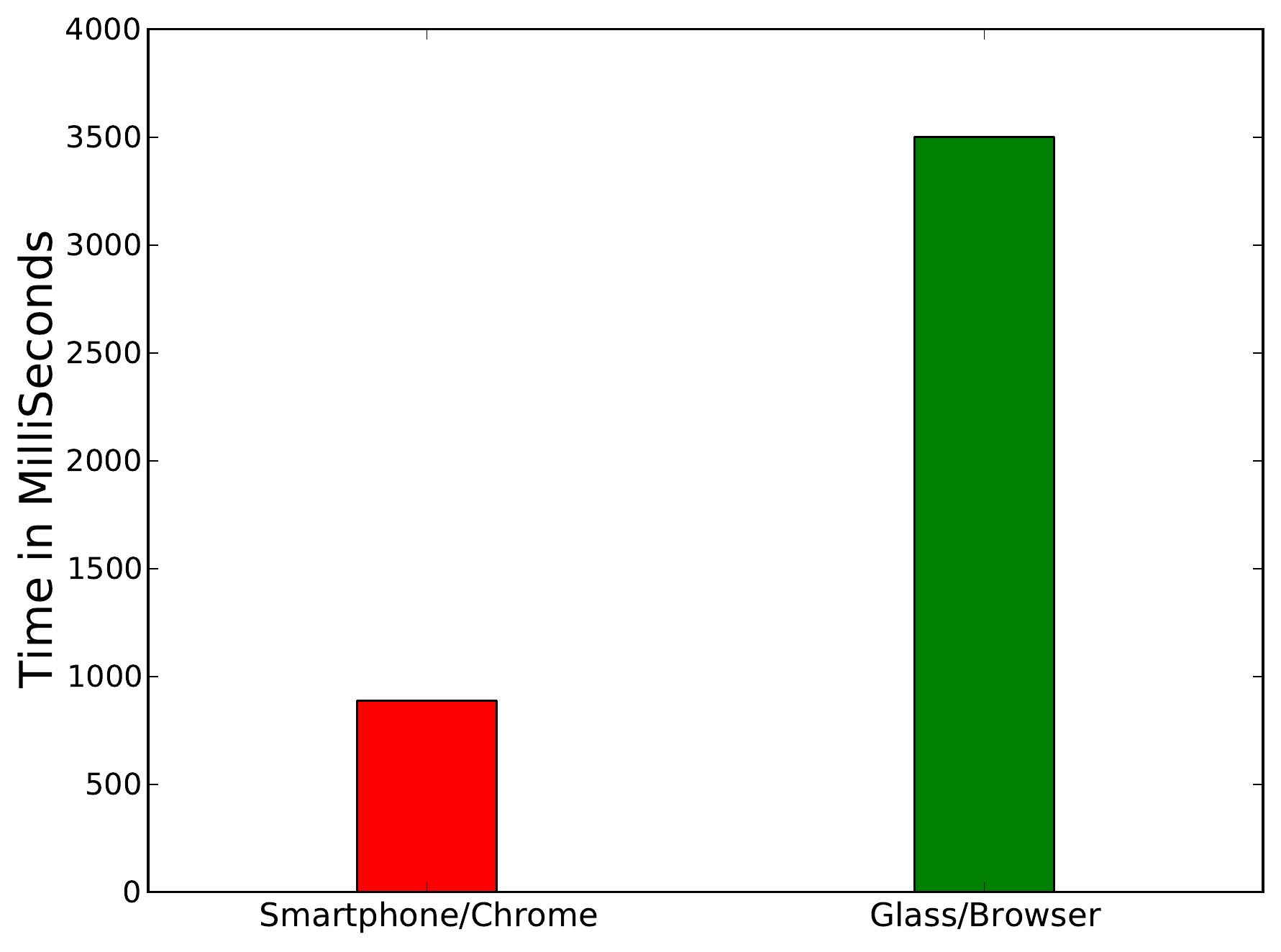}
\caption{ SunSpider  Benchmark}
\label{fig-sunspider}
\end{subfigure}%
    \begin{subfigure}{0.7\figwidth}

  \includegraphics[width=\textwidth]{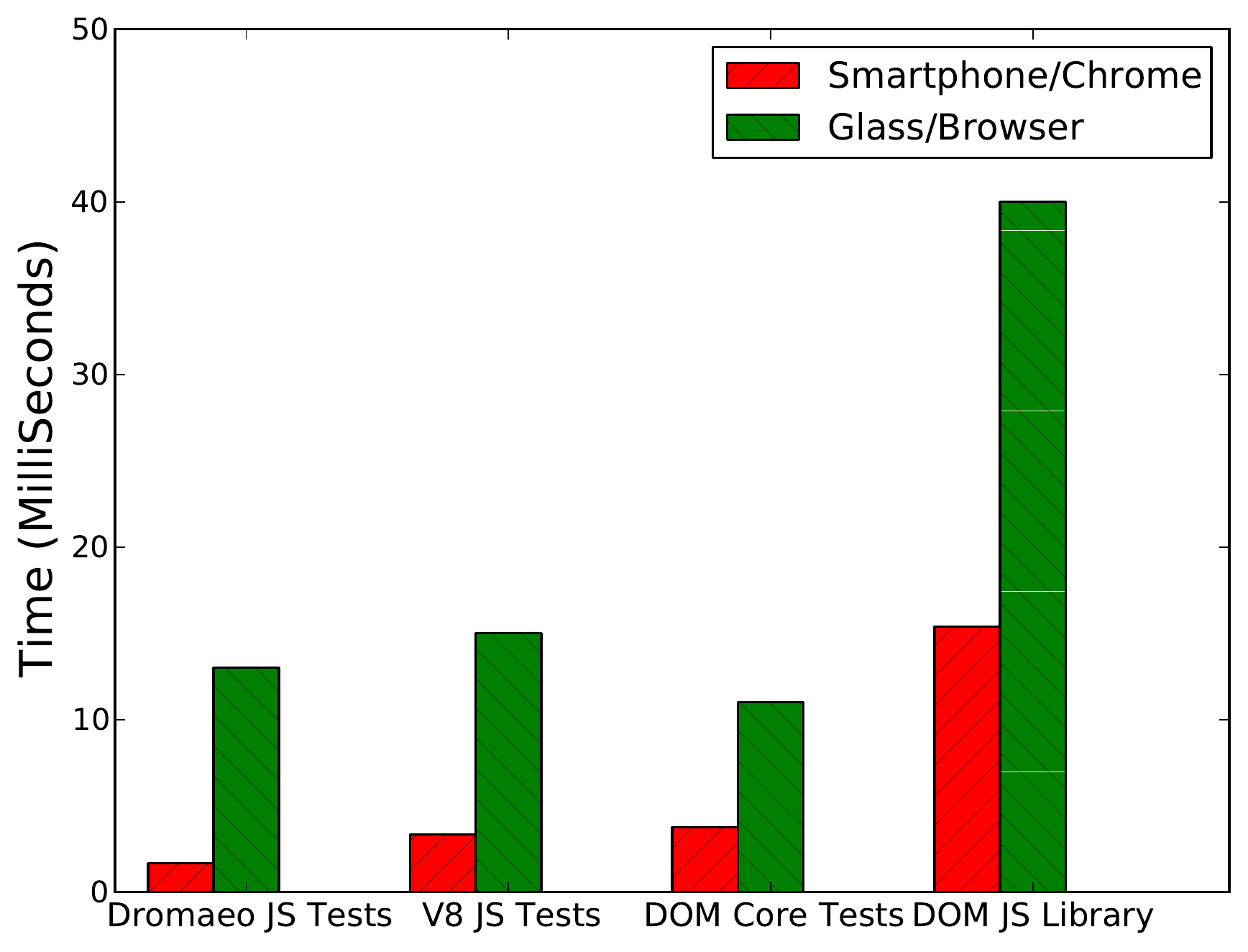}
\caption{ Dromaeo  Benchmark}
\label{fig-dora}
\end{subfigure}%
\begin{subfigure}{0.7\figwidth}
  \includegraphics[width=\textwidth]{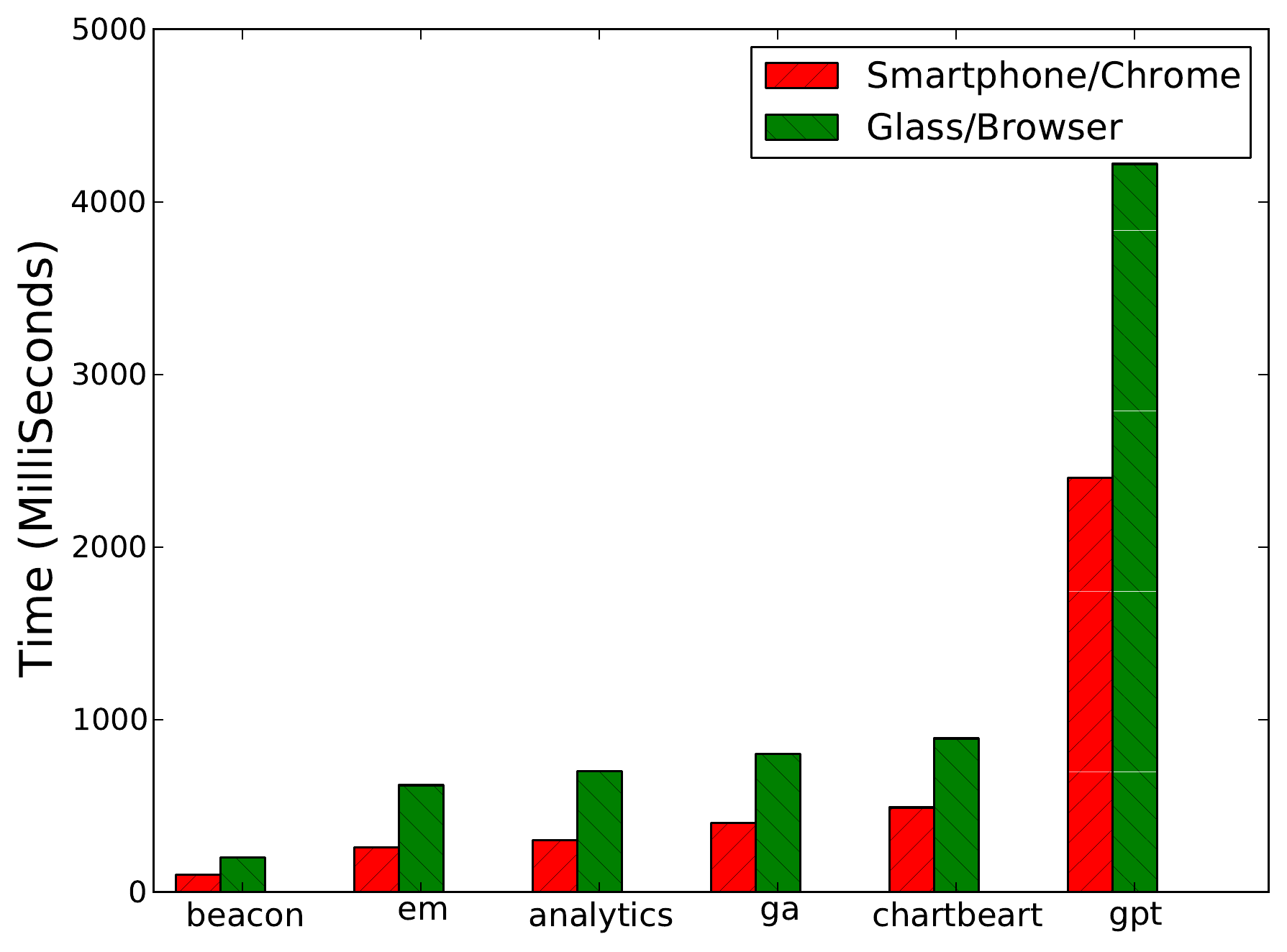}
\caption{Ads and Analytics}
\label{fig-ads}
\end{subfigure}
\caption{Time to execute  JavaScripts}
    \label{fig-javascripts}
\end{figure*}


The results  are shown in Figure \ref{fig-syn}.  In general, JavaScript is the most power hungry web component across  the websites as shown in  Figure \ref{fig-webcomp}. JavaScript  consumes more than 40 \% of the total power  on 10 of the 12 websites.   JavaScript is also the highest contributor to the webpage load time  as shown in Figure \ref{fig-jstiming}. The presence of JavaScript content on  today's websites is very recurrent.  A recent study \cite{Mendoza} indicates that 33 \% of the total JavaScript on mobile webpages remain unused on a smartphone  browser. Analytics, Ads and tracking-related scripts constitute a high number of JavaScripts that is not critical to the functional processing of the webpage and is of limited utility to a user. Special attention should be given to the treatment and inclusion of JavaScripts on  small devices browsers.

\subsubsection{JavaScript Benchmarks: Comparison}

As JavaScript is the most resource intensive component,  we   compared the JavaScript execution time across some popular benchmarks: SunSpider version 1.02 \cite{sun}, Dromaeo suite \cite{dromaeo}, and  ads and analytics scripts: beacon.js, em.js, analytics.js, ga.js, chartbeat.js and gpt.js   on Glass and smartphone. This provides an estimate as to how slow Glass is compared to  a smartphone when running the same JavaScript. We executed all 26 JavaScripts from SunSpider. From Dromaeo suite, Dromaeo JavaScript Tests, V8 JavaScript Tests, DOM Core Tests and JavaScript Library Tests were executed. The results are shown in Figure \ref{fig-javascripts}.






  Sunspider benchmark on Glass/Browser is  4x slower than Smartphone/Chrome  as shown in Figure \ref{fig-sunspider}. The  execution time  for  Dromaeo  suite  in Figure \ref{fig-dora} show that the Glass/Browser is about 3-8 times  slower than Smartphone/Chrome browser,  while executing the same JavaScript. These results  again highlight the need for  today's websites to serve less JavaScript to Glass. The results    in Figure \ref{fig-ads} shows that the  execution time for 3rd party scripts on Glass is about 2x that of smartphone. Google Publishing Tag (gpt.js)  is the most time consuming script, which suggests that serving ads on Glass  can cause significant  delay in loading webpages. 





\subsubsection{Image Formats: Comparison}

\begin{figure}[!h]
  \centering
  \includegraphics[width=0.4\textwidth]{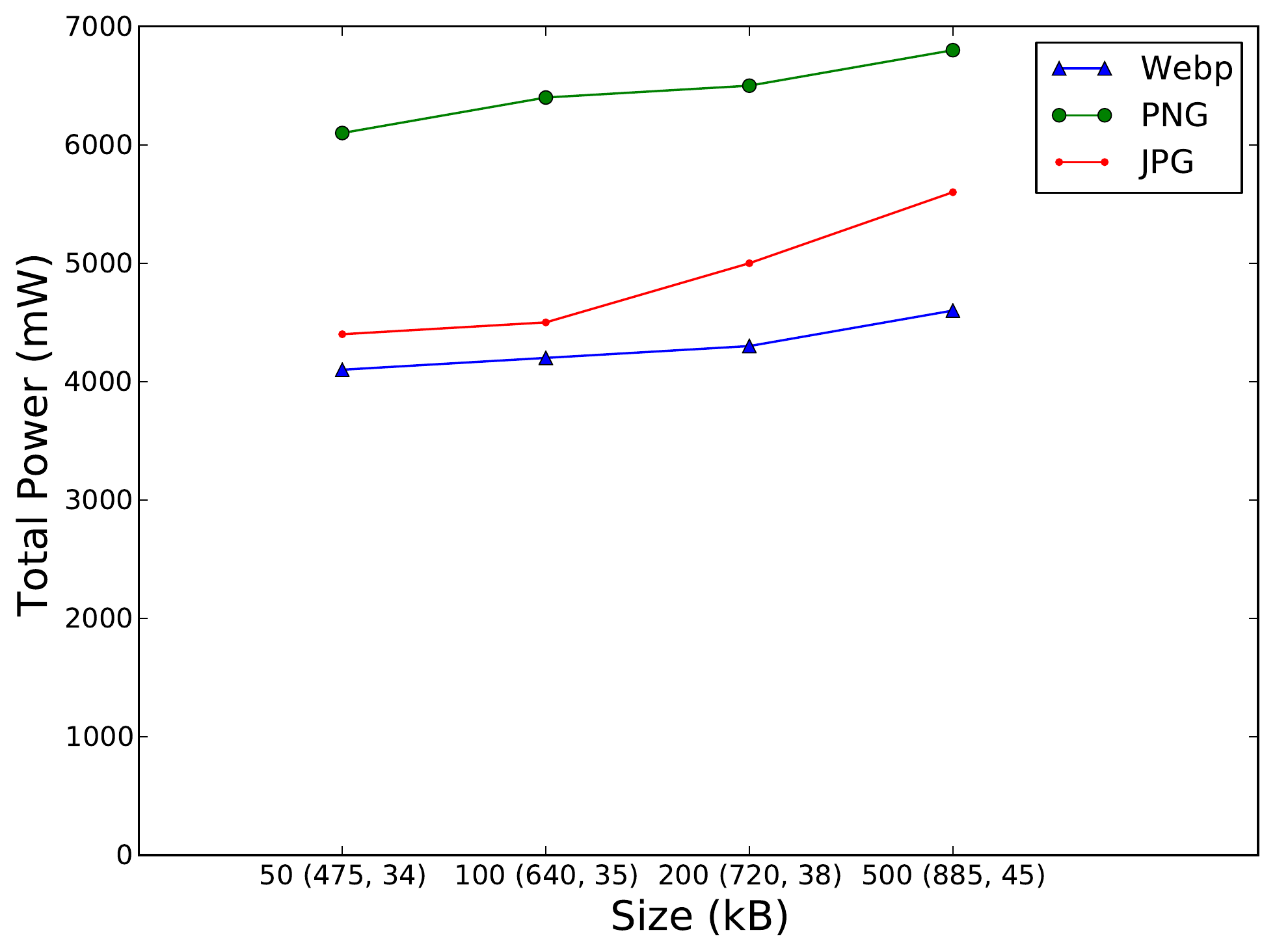}
\caption{Power consumption for image formats on Glass}
\label{fig-images}
\end{figure}

We measured the energy consumption of PNG, JPEG and WebP  image formats on Glass. A JPEG image of size 500 KB is chosen for the experiments.  This JPEG image is then converted to smaller JPEG images and similar  PNG and WebP images using cwebp  \cite{cwebp}.  Each image is then embedded in a webpage (that contains only the image) on  local Apache web server.  Figure \ref{fig-images} shows the results. The x-axis shows firstly the JPEG image size. The two numbers inside the bracket represent the corresponding size for PNG and WebP image. Study by Josh et al. \cite{josh} showed that  depending on the quality comparison algorithm, the compression  ratio between JPG and WebP image files can be less than or greater than one while  maintaining  the same  quality of image at a particular JPEG level. On contrary, we  do not exclusively focus  on comparing the quality of the image.  We choose a particular JPEG file size which is  converted to a WebP image with a quality factor of 75 \%.  

\begin{figure*}[ht]
\centering
\begin{subfigure}{\figwidth}
\includegraphics[height=0.25\textheight]{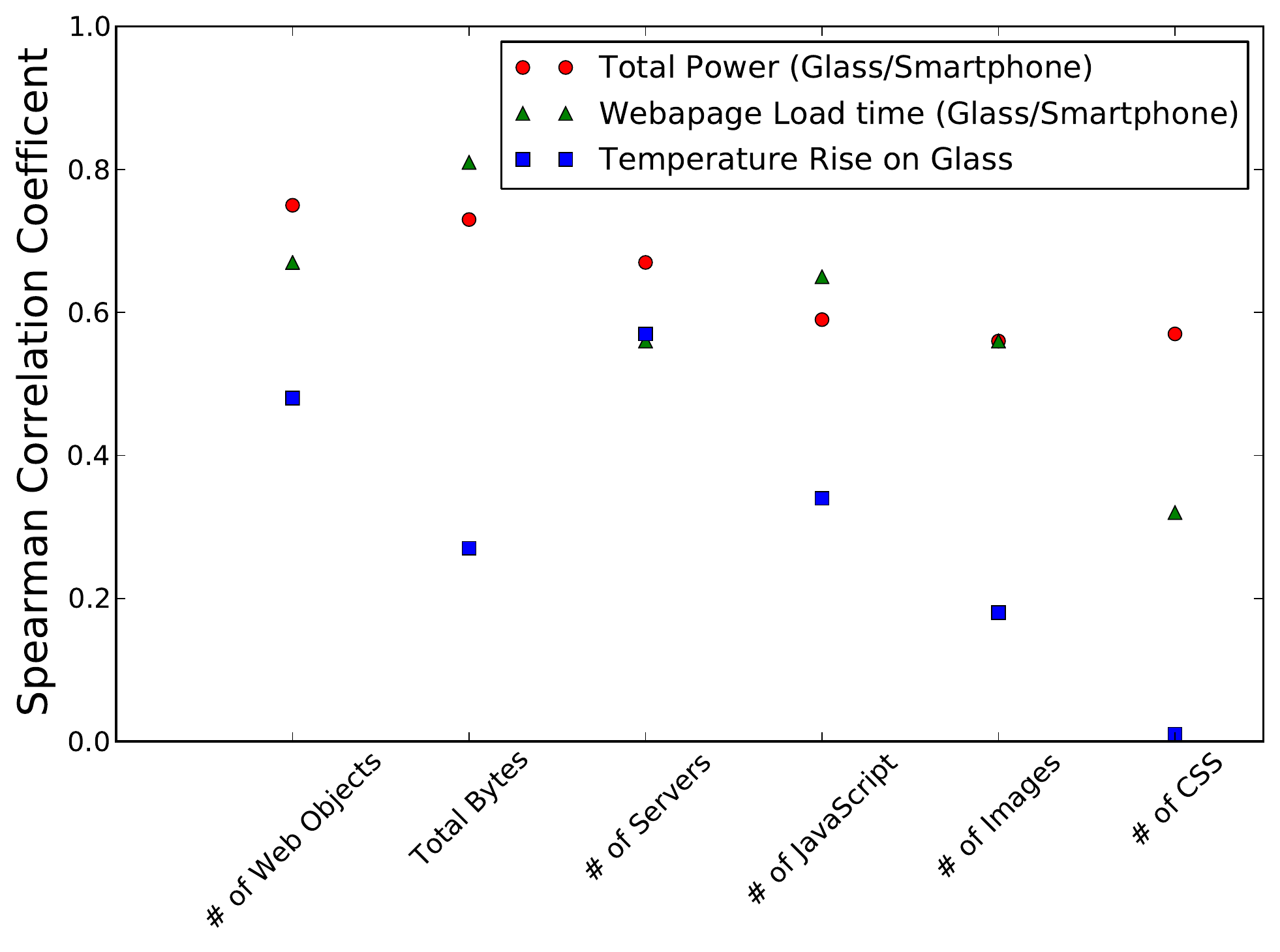}
\caption{Correlation Analysis (HTTPS/HTTP)}
\label{fig-corrfighttps}
\end{subfigure}%
\begin{subfigure}{\figwidth}
     \includegraphics[height=0.25\textheight]{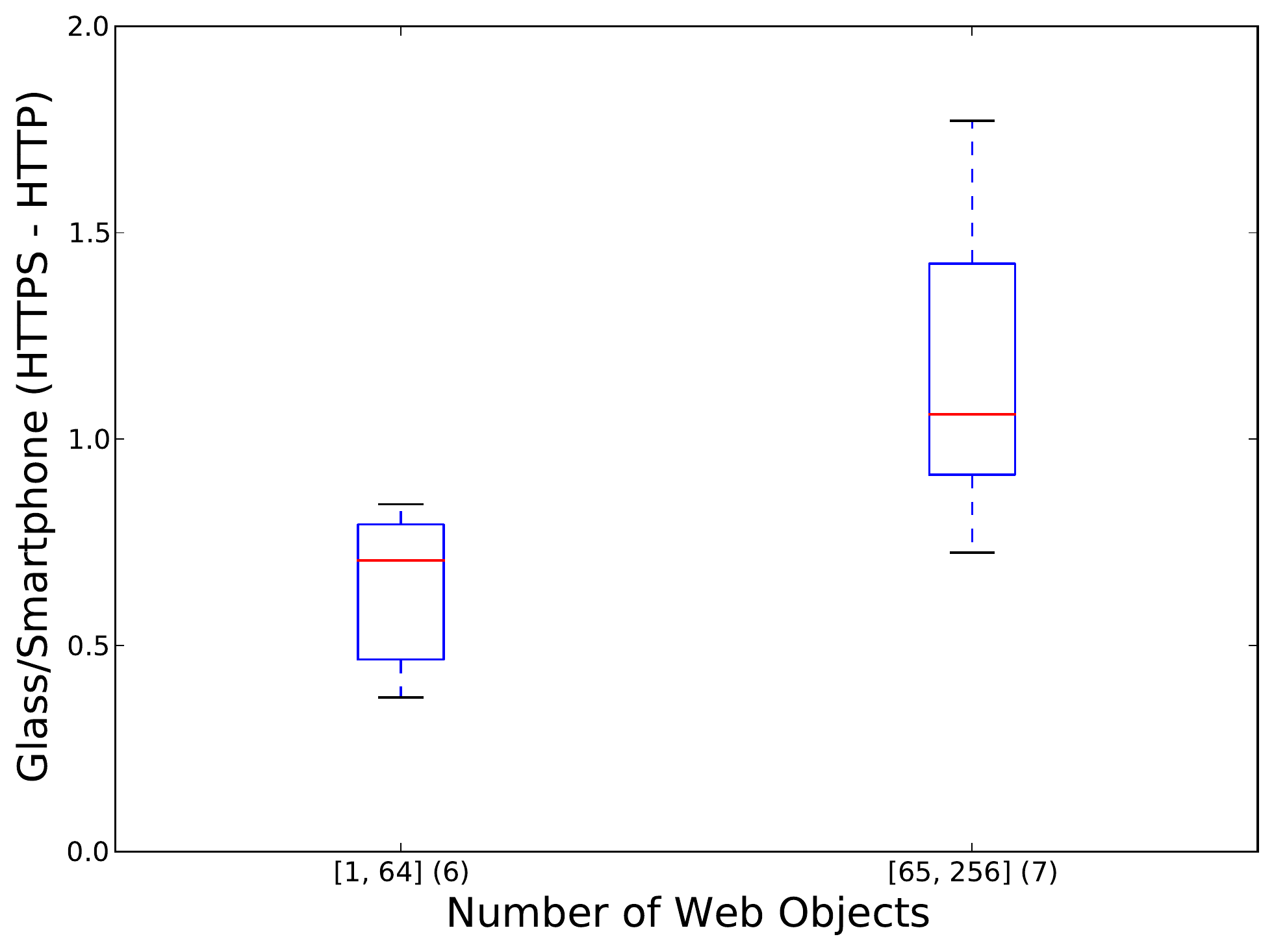}
     \caption{Total Power Consumption (mW)}
     \label{fig-boxplothttps1}
\end{subfigure}
\begin{subfigure}{\figwidth}
	\includegraphics[height=0.25\textheight]{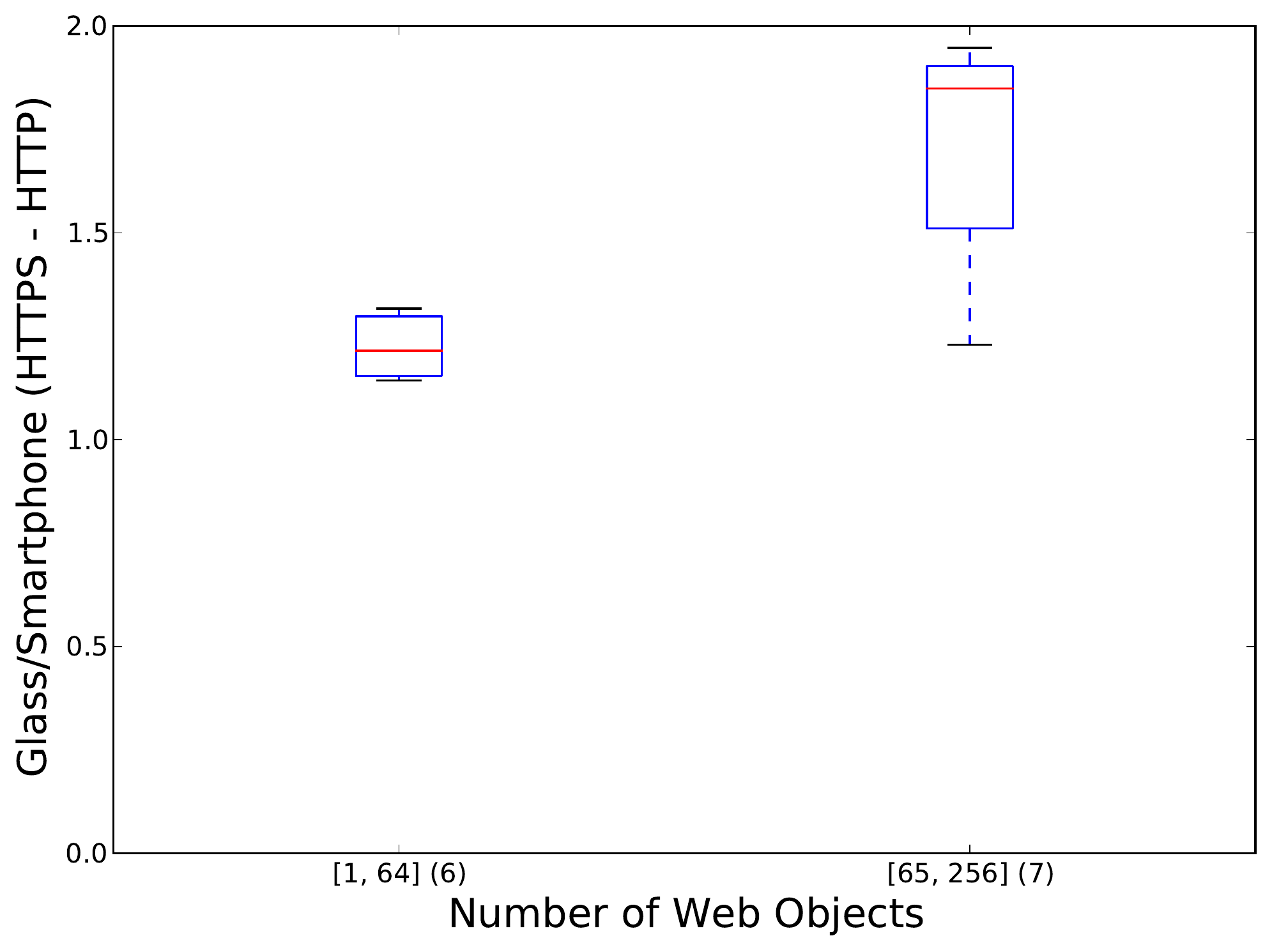}
    \caption{Webpage Load Time}
    \label{fig-boxplothttps2}
\end{subfigure}%
\begin{subfigure}{\figwidth}
     \includegraphics[height=0.25\textheight]{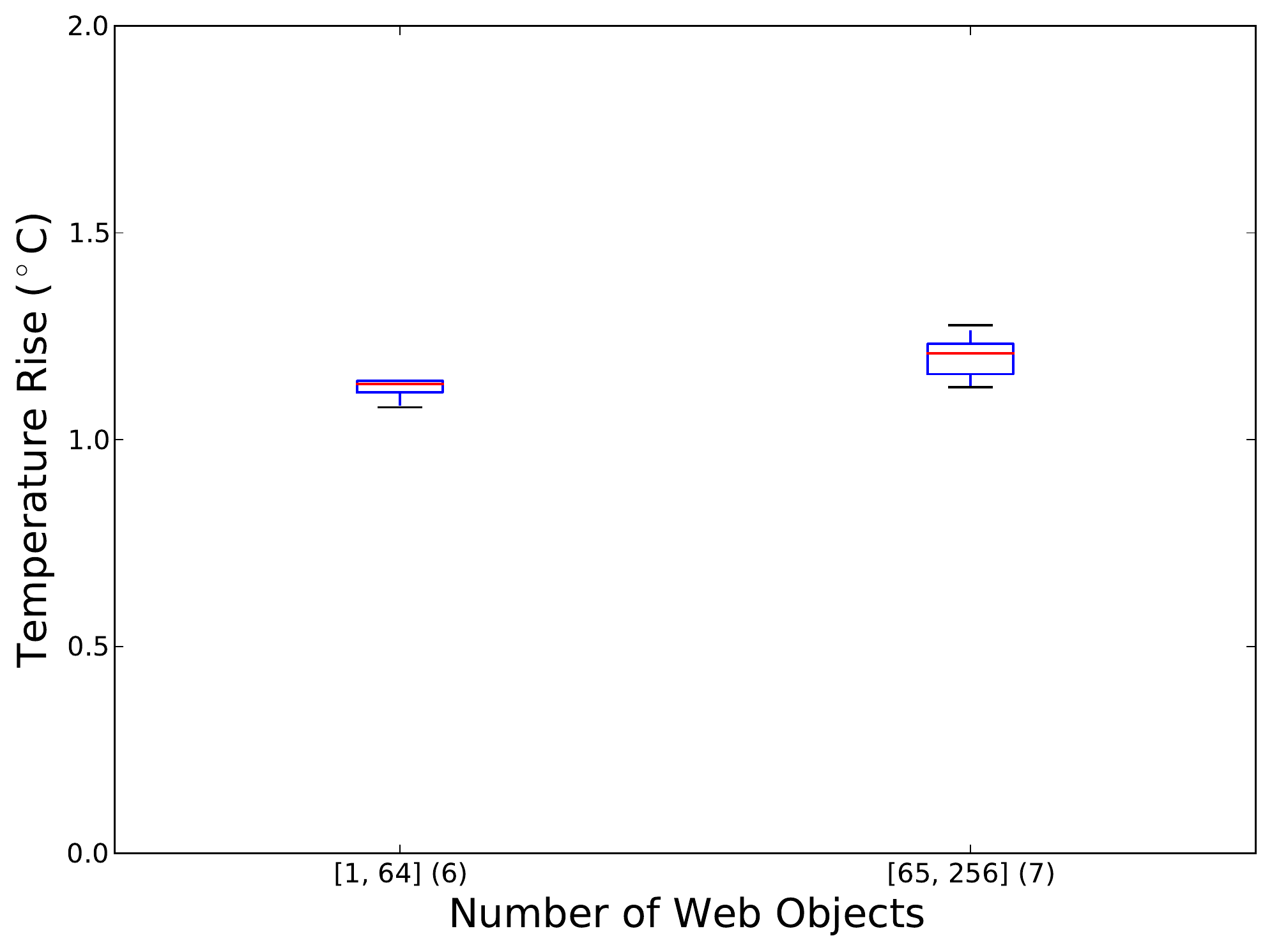}
     \caption{Temperature Rise on Glass}
     \label{fig-boxplothttps3}
     \end{subfigure}
\caption{Comparison of HTTP vs HTTPS Performance}
\label{fig-https}
\end{figure*}
 
Our results suggest that the performance gap between JPEG and WebP increases with increasing image file size because at higher image sizes, WebP gives a better compression ratio that results in smaller  WebP files and hence is the most energy efficient format.  Hence, webpages can be embedded with WebP format instead of PNG or JPEG to achieve lower power consumption on under powered  devices. As an example,  converting all JPEG and PNG  images to WebP   provides savings of 20 \% in power consumption and 33 \% lower  webpage load time for \url{ted.com}. Similar conversion on \url{m.wikihow.com} gives 45 \% savings in power and 50 \% lower   webpage load time.




\subsection{Cost of HTTPS}

We measured the performance of HTTPS versus HTTP on Glass and compare it to a smartphone.  Webpage load time is measured while accessing  a website using HTTP and HTTPS  and  power consumption, temperature variation and downloaded  bytes is calculated  for the duration of the webpage load time. 18  out of 50 websites supported both HTTP and HTTPS.  However, we could only compare results for 13 websites because two website loads less content on HTTPS than HTTP and three websites have been optimized for Glass. Similar to Section 4.1,  relative performance (Glass/Smartphone) ratio is used to represent the results. 



We did a correlation analysis similar to  Section 4.1 and then binned websites based on the common highest correlated factor to depict the relative performance. The result is shown in Figure \ref{fig-https}. Figure  \ref{fig-corrfighttps} shows that   the number of web objects  as the common factor among the top three factors for the three performance metrics: relative total power consumption, relative webpage load time, and temperature variation on Glass. 

 Figure \ref{fig-boxplothttps1}-\ref{fig-boxplothttps3} shows that the relative performance of Glass compared to a smartphone deteriorates with increasing number of web objects on a webpage.  For the webpages with smaller number of web objects (<64), Glass webpage load time is 27 \%  higher than smartphone. However, Glass power consumption is 27 \% lower than smartphone due to the lower rate of power consumption.  With increased number of web objects (>64), we see   69 \% higher webpage load time and 17 \% higher power consumption on Glass compared to a smartphone.    The  cost of HTTPS  can be attributed to the extra time to maintain and create HTTPS connections which  increases  with increasing number of  web objects  and the number of  servers to contact. More time also leads to  more power consumption and  higher  temperature on Glass.


\subsection{Optimized Websites}
Seven websites: \url{m.espn.go.com}, \url{m.wikipedia.com},  \url{m.wikihow.com}, \url{goodreads.com}, \url{telegraph.co.uk}, \url{m.youtube.com}, and \url{mobile.bloomberg.com} provide tailored content  to Glass. The general methodology employed by these tailored websites is to deliver less content to Glass than smartphone.  Next, we discuss specific features discovered by analyzing tcpdumps. 

  \url{m.espn.go.com} and \url{m.wikipedia.com}  optimize by delivering the images to Glass  browser according to the device  screen dimensions which results in 50 \% and 70 \% reduction in image content download respectively.   Considering the number of images on \url{m.espn.go.com} (35) and \url{m.wikipedia.com} (26), the reduction is considerable.   Glass screen dimension (427*240) is smaller than Nexus 5 (640*360). Upon receiving request  from any device, the  server identifies the device type and its display properties,
which is then used to send appropriate sized content to the device. Accessing  \url{m.wikipedia.com}  and \url{m.wikihow.com} on Glass fetches less php scripts than smartphone, resulting in savings of (450 kB).  \url{goodreads.com} does not fetch a particular CSS on Glass which is required only for smartphone and hence saves 700 kB of traffic. \url{telegraph.co.uk}  optimize by not  showing ads on Glass and  thereby avoiding 1 MB of ad related scripts.

\url{m.youtube.com} and \url{mobile.bloomberg.com} have a different version of website for Glass and smartphone. Note that  different version  mean accessing \url{m.youtube.com} or \url{mobile.bloomberg.com} fetches a different HTML file for the landing page.  \url{m.youtube.com}  serves 50 \% (700 kB) lesser  and \url{mobile.bloomberg.com} serves 60 \% (1.5 MB) lesser content to Glass than smartphone version of the website.

\section{Conclusion}
In general, the browsing  performance  on Glass is worse than  a smartphone primarily   because the same content is being delivered to Glass and smartphone regardless of the device type. Glass is an underpowered mobile device with smaller processing power  and  smaller battery than smartphones. The following suggestions might be beneficial for efficient  browsing on underpowered devices: 
\begin{itemize}
\item Reducing web content: As JavaScripts significantly impede the   performance of  Glass browser, their  quantity and complexity can be reduced.  Serving less CSS, images, ads and keeping in consideration the display capability of the requesting device can also reduce the content. 
\item Efficient image formats: WebP  instead of JPEG or PNG can be used on webpages to  save power and lower the webpage load time.






\item Cloud assisted solutions: An underpowered device can have a light weight browser using cloud based acceleration rather than a full fledged browser as  Glass possess. Another way is    using a data compression proxy  on the cloud, thereby reducing network bandwidth, power consumption and webpage load time.
\item New protocols: SPDY can be used in place of HTTPS to improve  browser performance.

\end{itemize}

\bibliography{sigproc} 
\bibliographystyle{plain}
%
%

%
%
\appendix

\begin{table} [h]  
\centering
 \begin{tabular}{|>{\centering\arraybackslash}p{8cm}|}  \hline
  Website \\\hline
  \url{m.espn.go.com}, \url{m.wikipedia.com},  \url{m.wikihow.com}, \url{goodreads.com}, \url{telegraph.co.uk}, \url{m.youtube.com}, \url{mobile.bloomberg.com}, \url{bing.com}, \url{quora.com}, \url{m.rediff.com}, \url{java.com}, \url{i.reditt.com}, \url{tripadvisor.com}, \url{apple.com}, \url{m.yelp.com}, \url{booking.com}, \url{go.com}, \url{ted.com}, \url{m.9gag.com}, \url{m.aol.com}, \url{adobe.com}, \url{deviantart.com}, \url{gizmodo.com}, \url{skype.com}, \url{m.gsmarena.com}, \url{nhl.com}, \url{dell.com}, \url{taboola.com}, \url{m.goal.com}, \url{m.bleacherreport.com}, \url{m.indiatimes.com}, \url{diply.com}, \url{m.huffpost.com}, \url{microsoft.com}, \url{m.foxnews.com}, \url{m.bbc.com}, \url{m.ndtv.com}, \url{theladbible.com}, \url{techcrunch.com}, \url{businessinsider.com}, \url{m.imdb.com}, \url{m.mlb.com}, \url{engadget.com}, \url{answers.com}, \url{photobucket.com}, \url{theguardian.com}, \url{mweb.cbssports.com}, \url{lenovo.com}, \url{dailymail.com}, \url{m.espncricinfo.com}  \\\hline
\end{tabular}
\caption{Websites Used In Measurements}
\label{table-redirections}
\end{table}
\balancecolumns
\end{document}